\documentclass[aps,pre,longbibliography,lengthcheck,a4,floatfix]{revtex4}
\usepackage[shortlabels]{enumitem}
\usepackage{graphicx}
\usepackage[version=3]{mhchem}

\usepackage{xcolor}
\begin{document}
\title{From ionic surfactants to Nafion through convolutional neural networks}
\author{Lo\"ic Dumortier}
\affiliation{X-ray Microspectroscopy and Imaging Group, Department of Chemistry,
 Ghent University, Krijgslaan 281/S12, B-9000 Gent, Belgium}
\author{Stefano Mossa}
\affiliation{Univ. Grenoble Alpes, CEA, IRIG-MEM, 38000 Grenoble, France}
\affiliation{Institut Laue-Langevin, BP 156, F-38042 Grenoble Cedex 9, France}
\email{stefano.mossa@cea.fr}
\date{\today}
\begin{abstract}
We have applied recent machine learning advances, deep convolutional neural network, to three-dimensional (voxels) soft matter data, generated by Molecular Dynamics computer simulation. We have focused on the structural and phase properties of a coarse-grained model of hydrated ionic surfactants. We have trained a classifier able to automatically detect the water quantity absorbed in the system, therefore associating to each hydration level the corresponding most representative nano-structure. Based on the notion of transfer learning, we have next applied the same network to the related polymeric ionomer Nafion, and have extracted a measure of the similarity of these configurations with those above. We demonstrate that on this basis it is possible to express the static structure factor of the polymer at fixed hydration level as a superposition of those of the surfactants at multiple water contents. We suggest that such a procedure can provide a useful, agnostic, data-driven, precise description of the multi-scale structure of disordered materials, without resorting to any a-priori model picture.
\end{abstract}
\maketitle
\section{Introduction}
\label{sect:intro}
Nafion is the reference material employed for the membrane in polymer electrolyte membrane fuel cells (PEMFC)~\cite{kreuer2013ion}. Once hydrated, this ionomer organizes in a disordered phase-separated structure, where a strongly hydrophobic matrix, providing the mechanical strength, is separated from extended ionic domains by charged interfaces. This structure is strongly dependent on the hydration level, often expressed by the parameter $\lambda$, {\em i.~e.}, the number of absorbed water molecules per sulfonic acid group terminating the side chains. {\color{black}Optimization of FC performances implies a thorough understanding of the impact of this complicated multi-scale organization on the correlated transport of charge carriers (protons) and water molecules.}

Insight on Nafion nano-morphology mainly comes from small-angle neutron and X-ray scattering. These are powerful tools which, however, provide strongly space-averaged information, as typical of reciprocal space techniques. Analysis of the static structure factor, $S(Q)$, only allows to extract estimates of the average size of the ionic domains, from the position of the ionomer peak (see, for instance,~\cite{gebel2000structural}). Any attempt to develop a picture of the 3-dim nano-morphology therefore finally rests on a wise choice of stylized arrangements of simple geometry domains. This is an intrinsically arbitrary procedure, with the only constraint of consistency with the scattering data and, possibly, main physicochemical principles~\cite{kreuer2013critical}. 

While over the years a few competing alternatives have been proposed, ranging from spherical domains~\cite{hsu1983ion} through fibrillar (bundle) structures~\cite{rubatat2004fibrillar}, to parallel cylinders arrangements~\cite{schmidt2008parallel}, no final consensus has been reached. {\color{black}Even computer simulations, ranging from all-atoms~\cite{venkatnathan2007atomistic} and coarse-grained~\cite{allahyarov2011simulation} (CG) Molecular Dynamics (MD) to dissipative particle dynamics (DPD)~\cite{vishnyakov2014self}, have not been of great help to solve the uncertainties. Reaching length scales and spatial resolution sufficient to fully characterize properties and behavior of the proposed morphological models remains, in fact, extremely difficult~\cite{knox2010probing}.} Pitfalls and limits of this method being evident~\cite{kreuer2013critical}, model-agnostic approaches, not bounded to a particular prior for morphology, would clearly be preferable. These include from advanced statistical analysis, like the Maximum Entropy study of small-angle scattering and mesoscopic simulations data of~\cite{elliott2011unified}, to direct real-space imaging, like the cryogenic electron tomography of~\cite{allen2015morphology}, which avoids all-together the difficulties coming from the lack of phase information.  

Somehow mediating between the above perspectives, it has been recently realized that, at the nano-scale, Nafion structure and transport therein are similar to those observed in related but less elusive materials, sulfonated ionic surfactants~\cite{lyonnard2010perfluorinated,hanot2015water,hanot2016sub,Berrod2017}. These are macro-molecules very similar to the Nafion side chain but, in contrast with the ionomer, their phase modifications with hydration are characterized precisely: lamellar, hexagonal, and micellar phases appear one after the other on increasing $\lambda$~\cite{lyonnard2010perfluorinated}. In addition, it has been demonstrated that the average size of the ionic domains in Nafion, in the related Aquivion, and in ionic surfactants are very similar, in a quite large range of hydration~\cite{Berrod2017}. These observations make of surfactants good proxies for better grasping both nano-morphology and properties of the ionomer.
\begin{figure*}[t]
\centering
\includegraphics[width=0.99\textwidth]{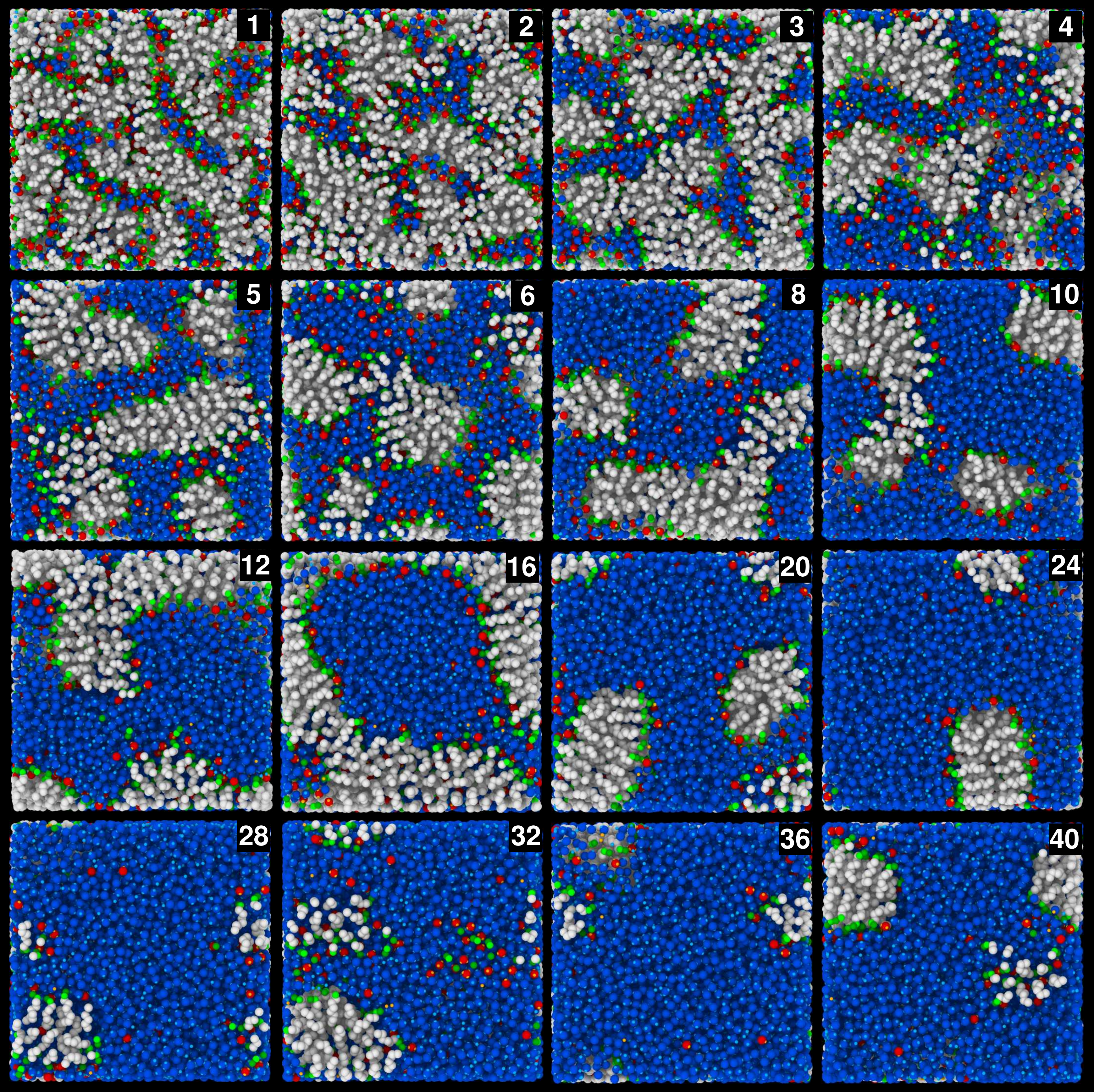}
\caption{Ionic surfactant MD computer simulation snapshots, at the indicated values of hydration, $\lambda_S$. The hydrophobic sections of the surfactants are in white, the polar head in green. The adsorbed water molecules are depicted in blue, the hydronium ion complexes in red. One can clearly appreciate the modifications from lamellar phases at low $\lambda_S$, through increasingly elongated structures, to micelles at the highest values of water content. Some of these structures have been reported in~\cite{hanot2015water}. These data are discussed at length in the main text.}
\label{fig:surfactant_snaps}
\end{figure*}

Here we attempt to make the above picture more precise and, developing on the same line, we explore the possibility that a chunk of Nafion at a given hydration $\lambda_N$ could be meaningfully described as a disordered collection (tiling) of patches corresponding to typical ionic surfactants morphologies, possibly at multiple hydrations $\lambda_S$. To develop this program, we have employed a mapping procedure which does not resort to any a-priori view of the materials nanostructure, based on recent Machine Learning (ML) techniques. 

ML methods, in particular Deep Learning (DL)~\cite{lecun2015deep}, are increasingly integrating the palette of numerical tools employed in modern science, ranging from materials informatics~\cite{rickman2019materials} to fundamental physics~\cite{carleo2019machine,ferguson2017machine}. We are interested in the application of DL to images-related tasks~\cite{kalinin2015big}, with emphasis on 3-dimensional (tomography) data sets. A most relevant technology in this context is Convolutional Neural Networks (CNN)~\cite{lecun2015deep}, which have been demonstrated to be an extremely powerful tool in many different fields. 

In what follows we detail step-by-step the framework we have developed in this context, integrating different computational techniques. We have generated by Molecular Dynamics (MD) computer simulation extensive data bases of ionic surfactants and Nafion configurations, in very large hydration ranges. These data, opportunely treated, have been employed for training and inference of a 3-dimensional convolutional neural network. We show how it is possible to use the CNN outputs to gain additional insight on the nanomorphology of the ionomer, also in terms of the static structure factors. We conclude with a discussion of the implications of our findings and possible perspectives of similar approaches. All details of the numerical methods are deferred to the Methods section.
\section{The surfactant phases classifier}
\label{sect:workflow}
We have implemented a supervised-learning 3-dimensional configurations classifier able to identify ionic surfactant configurations generated at different water content values $\lambda_S$ (the labels). This first step provides us with an automatic tool that, when presented with an ionic surfactant configuration (not comprised in the training set), outputs the water content therein together with the associated likelihood, therefore unambiguously identifying the corresponding phase state. Although similar goals have already been achieved for recognizing, for instance, the symmetry of crystal domains in nano-structured materials~\cite{vasudevan2018mapping}, this is to the best of our knowledge, a first in a soft matter context. This is also the workhorse we exploit for the additional step forward that follows. Our workflow is organized as described below.
\subsection{Data}
\label{subsect:data} 
We have generated a comprehensive training data set including 16 $\lambda_S$ values in the range $0\le\lambda_S\le 40$, by massive MD simulation of the coarse-grained (CG) model of~\cite{hanot2015water} and references therein. We have used LAMMPS~\cite{plimpton1995fast}, a High Performance Computing tool for MD simulation. After thermalization, we have produced, at temperature $T=300$~K and pressure $P=1$~atm in the $(NPT$)-ensemble, $10^3$ independent configurations at each $\lambda_S$, along a trajectory of $20$~ns. This procedure has been executed twice, initializing the systems from completely independent high temperatures realizations. This amounts to a total of $2\times 10^3$ samples at each $\lambda_S$, a very large number of system instances needed for the subsequent CNN training. All details of the MD computations are given in the Methods section.

We show typical system snapshots in Fig.~\ref{fig:surfactant_snaps}, at the indicated values of $\lambda_S$. With reference to the macromolecular structure detailed in~\cite{hanot2015water}, the strongly hydrophobic apolar sections of the surfactants are the white beads, while the negatively charged polar heads are constituted by two adjacent green beads. The ionic domains comprise the adsorbed water molecules displayed in blue, while the positively charged hydronium ions are in red. The phase behavior confirms the general features already described in~\cite{hanot2015water} for a more limited choice of $\lambda_S$ values. In particular, at the highest hydrations ($\lambda_S>20$) we recover a pure size-dispersed solution of micelles whose shape, for $12<\lambda_S<20$, gradually morphs from spherical to increasingly elongated structures. In the intermediate range $6<\lambda_S<12$, the aggregates start to merge and form extended bi-layers, transmuting at the lowest $\lambda_S<6$ into flat extended ionic domains. 

Note that, in general, the configurations at low hydrations look less ordered than those reported in~\cite{hanot2015water}, for mainly two reasons. First, due to the extremely large required configurations database, we have considered smaller system sizes than in that work. This implies smaller simulation boxes, which makes the macroscopic growth of the lamellar order more difficult. As we will see below, this amounts to the disappearance of the high-order Bragg peaks in the $S(Q)$ reported in~\cite{hanot2015water}, while the ionomer peak positions are not modified. Second, these systems have been thermalized and aged on exactly the same time scales needed to stabilize the Nafion morphology (see below), which were shorter than those of~\cite{hanot2015water}. This choice is motivated by the observation that Nafion morphology strongly evolves with the aging time~\cite{kreuer2013critical} and that is, therefore, important to probe structural features in the two materials on the same time scale. In any case, we have confirmed by visual inspection of several 3-dimensional configurations that flat extended ionic domains are indeed present locally, which is the important requirement here, as it will be clear in the following.   
\begin{figure*}[t]
\centering
\includegraphics[width=0.99\textwidth]{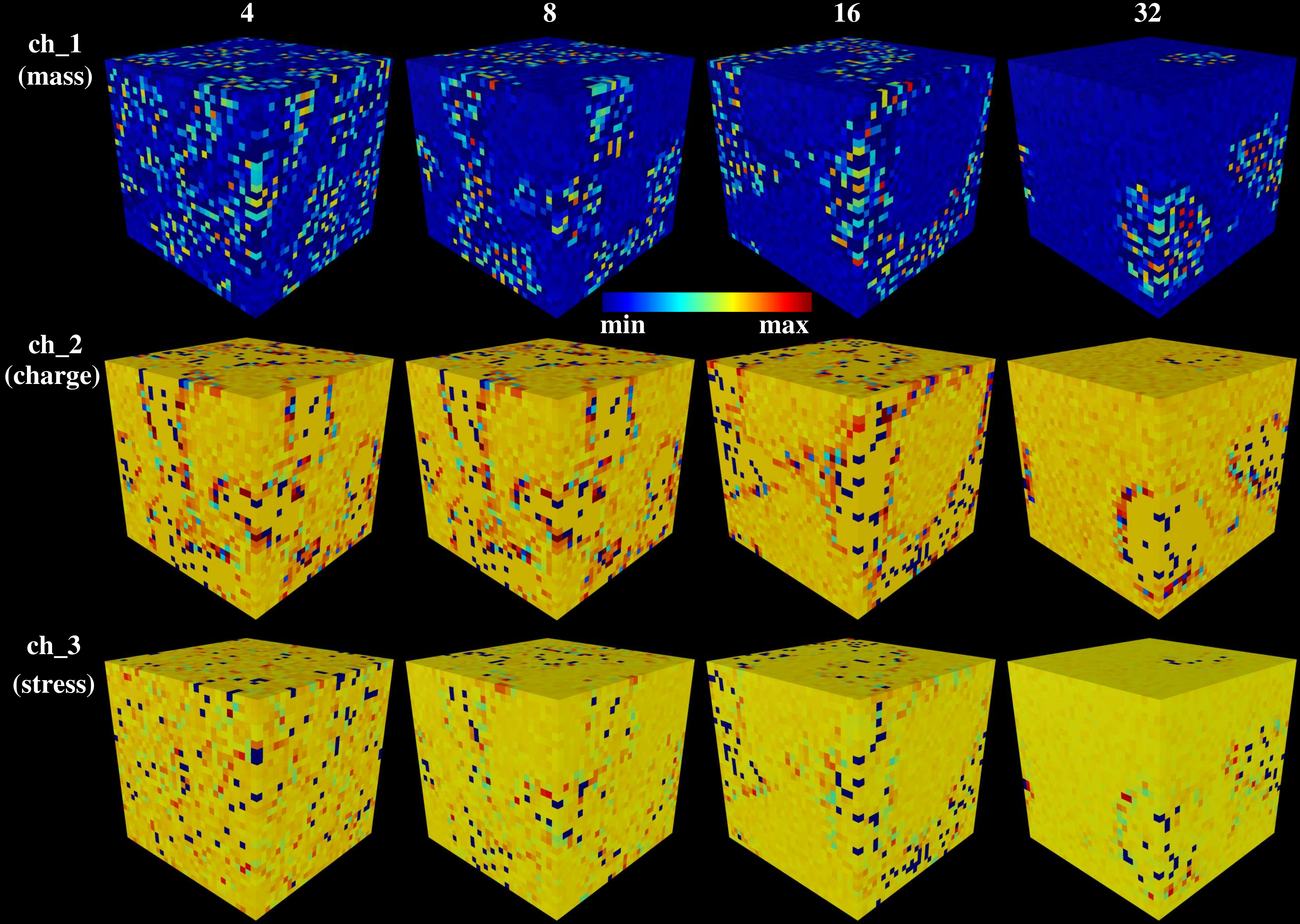}
\caption{From the ionic surfactant simulation snapshots to voxels with 3 channels, mass (1), charge (2), and local stress (3), at the indicated values of $\lambda_S$. Channel (1) provides comprehensive information about the mass distribution of the different moieties, channels (2) and (3) encode details of the charged interfaces separating the confining hydrophobic matrix from the structured ionic domains. These are the typical tomographic data (three-dimensional regular grids of size $M=32^3$) fed for training to the CNN implemented in Keras/TensorFlow~\cite{chollet2015keras}. Data preparation is described at length in the main text.}
\label{fig:voxels}
\end{figure*}

Next, we have encoded the simulated configurations with a 3-dimensional grid of voxels, each comprising three channels. These were chosen among the relevant system attributes. More in details, for each atom $a$ we have considered mass, $m^a$, charge, $q^a$, and off-diagonal components of the per-atom stress tensor, $\mathcal{S}_{\alpha\beta}^a=-m^a v^a_\alpha v^a_\beta-V_{\alpha\beta}$ (with $v_{\alpha,\beta}$ the $\alpha,\beta=x, y, z$ components of the atom velocity). Here, the first term is a kinetic energy contribution, the second is the virial term due to all intra- and inter-molecular interactions~\footnote{Note that the tensor is consistently defined in such a way that $-\sum_{a=1}^N\sum_\alpha\mathcal{S}_{\alpha\alpha}^a/3V=P$, the total system pressure.}. We have coarse-grained at run-time the above quantities on a regular grid of size $M=32^3$, averaging over the second half of $20$~ps time windows. In each voxel we have normalized the observables as $\Tilde{\mathcal{O}}=(\mathcal{O}-\mathcal{O}_\text{min})/(\mathcal{O}_\text{max}-\mathcal{O}_\text{min})$, where $\mathcal{O}_\text{min,max}$ are, respectively, the minimum and maximum values measured for each configuration.

While the above choice of the encoded features is arbitrary, it indeed provides a quite complete descriptor of morphology, as we demonstrate in Fig.~\ref{fig:voxels}, where we show typical voxels configurations at the indicated values of $\lambda_S$. Indeed, the mass channel (1) accurately describes the spatial distribution of the different moieties, therefore accounting for all information associated to phase separation of the apolar confining matrix from the ionic domains. The others embed relevant information about the nature of the interfaces, both at the level of interactions (2) and local mechanical response (3).
\begin{figure*}[t]
\centering
\includegraphics[width=0.99\textwidth]{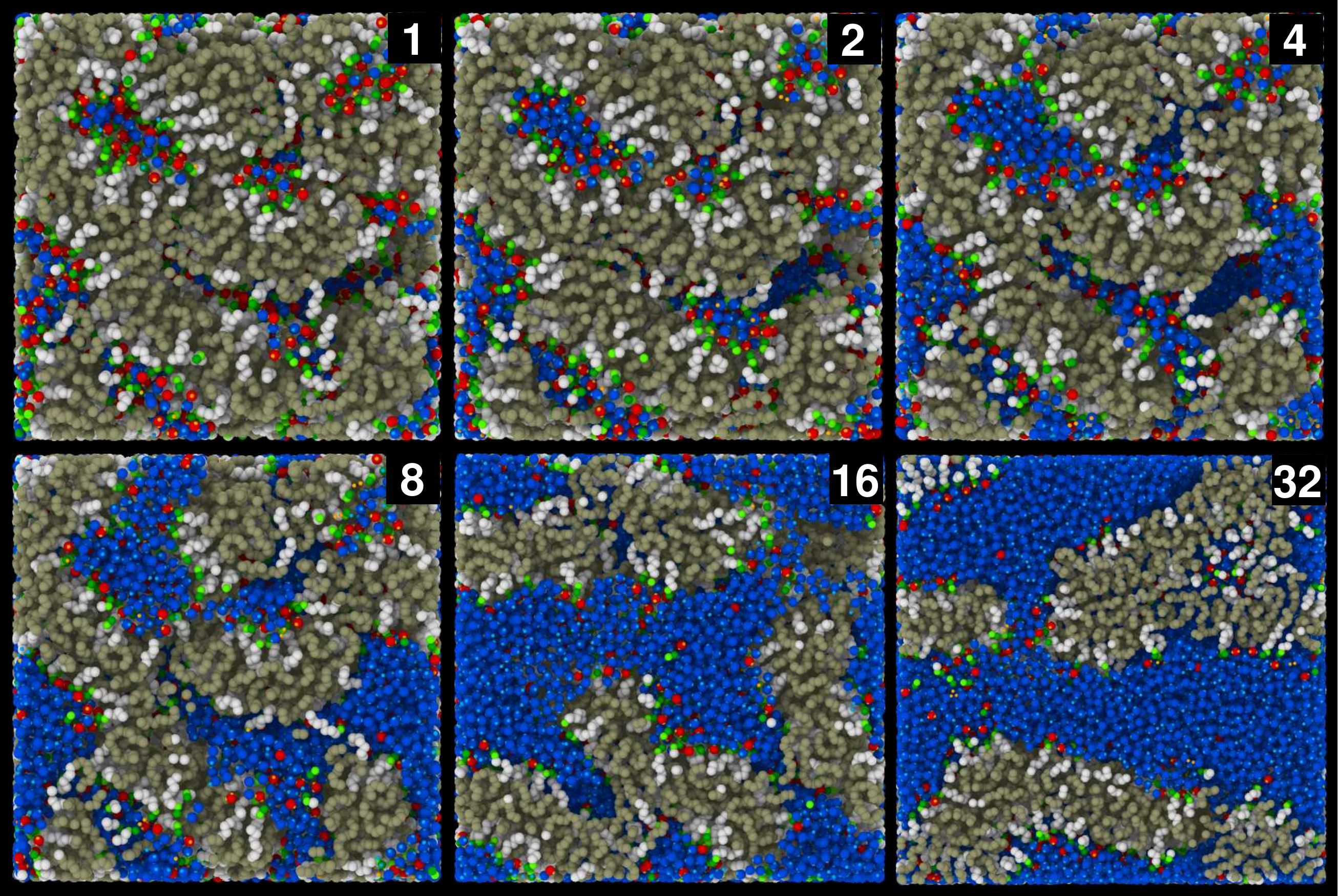}
\caption{Nafion MD snapshots at the indicated values of hydration $\lambda_N$. The color code is the same as in Fig.~\ref{fig:surfactant_snaps} for water molecules, hydronium ions and side chains, while hydrophobic backbones beads are depicted in grey. Some of these structures have been discussed in~\cite{hanot2016sub}.}
\label{fig:snap-nafion}
\end{figure*}
\subsection{Neural Network}
\label{subsect:model}
We have implemented a 3-dimensional CNN which we have trained against the above data sets. This particular choice seems the most natural for the problem, in order to keep the volumetric information on species distributions and grasp the hierarchy of structural motifs from the nano- to the meso- scale, by going deeper into the network layered structure. Indeed, the choice of a CNN is fully dictated by the nature of the system of interest. First, CNNs have been mostly developed to classify images and are therefore specialized in efficiently encoding the spatial structure of our voxel data. Note that, in principle, we could have made a different choice, partitioning the simulation box in 2-dimensional slices, therefore resorting to a more conventional 2-dimensional CNN. We have decided to keep the complete volumetric structure of the data, to maintain the possibility to fully describe inhomogeneities of the phase-separated domains in all directions. 

Second, a CNN is organized in such a way that, going forward in the deep architecture, each non-linear module transforms the representation at one level into another one at an increasing level of abstraction~\cite{lecun2015deep}. This allows to encode with comparable degree of accuracy an extended range of attributes, ranging from local properties like interfaces or mass distribution variations, to progressively more complex ones, like tortuosity or connectivity on growing length scales. We expect this process to grasp very naturally the intrinsic multi-scale nature of our systems morphology.    

All details of the 3-dimensional (volumetric) CNN are given in the Methods section. In general the choice of a particular architecture for the CNN is somehow arbitrary, and only trial and error allows to choose the structure most appropriate for a given problem. We describe our developments in the Methods section. For the implementation we have employed standard Python libraries including the Keras~\cite{chollet2015keras} high-level API to the TensorFlow~\cite{abadi2019tensorflow} machine leaning framework back-end. The training stage is depicted in Fig.~\ref{fig:cnn}, where we plot both accuracy and loss as a function of the number of epochs. 

The loss starts from a value $-\ln(1/16)\simeq 2.77$, expected for the initial random initialization of weights, reaching a very small value in about $30$ epochs. The extreme efficiency of the training stage is also confirmed by the fast increase of accuracy toward a value very close to $1$. Following training, we have tested the classification performances by inference on a data-set not included in the training set, comprising 3200 configurations, evenly distributed over all values of $\lambda_S$. Our network is eventually able to classify configurations in the entire hydration range, with an almost perfect accuracy of $99\%$. This level of precision is not surprising, due to the remarkable stability of the different phases (and therefore relatively limited variability of the generic features of the configurations) and the sheer size of the training sets. All-together, our results assure that we have available an extremely efficient classifier, able to discriminate among system morphologies as diverse as those shown in Fig.~\ref{fig:surfactant_snaps}.
\section{From ionic surfactants to Nafion}
\label{nafion}
We now illustrate our next step, that demonstrates the real benefit of the above classifier. The goal is to describe Nafion morphology in terms of the much better characterized ionic surfactant phases. We have therefore first generated by MD simulation, following the same procedure discussed above for the surfactants, an extended database of $7\times 10^3$ Nafion configurations~\cite{Berrod2017}, for 7 values of hydration, in the large range $0\le\lambda_N\le 32$ which also includes the $\lambda_N=0$ dry membrane condition. It is important to note that the model we have used for the ionomer side chains is exactly the same than that we have employed for the ionic surfactants. The chains are next grafted along the strongly hydrophobic polymeric backbone, with intramolecular interactions chosen to match a realistic value of the persistence length. The number of chains per polymer fixes the charge density of the ionomer. 

We show a few representative snapshots in Fig.~\ref{fig:snap-nafion}, at the indicated values of $\lambda_N$. Here the color code is the same that in Fig.~\ref{fig:surfactant_snaps}, with the difference that now the beads pertaining to the polymer backbones are depicted in gray. These data have been next prepared to be presented to the CNN exactly as described above for the ionic surfactants. Morphology indeed strongly changes with hydration, with well defined charged interfaces and the ionic domains modifying from thin, often disconnected, pores at very small $\lambda_N$, to large water pools at high $\lambda_N$. It is clear that it is not possible to safely associate any obvious nano-morphology to these disordered structures.  
\begin{figure}[t]
\centering
\includegraphics[width=0.49\textwidth]{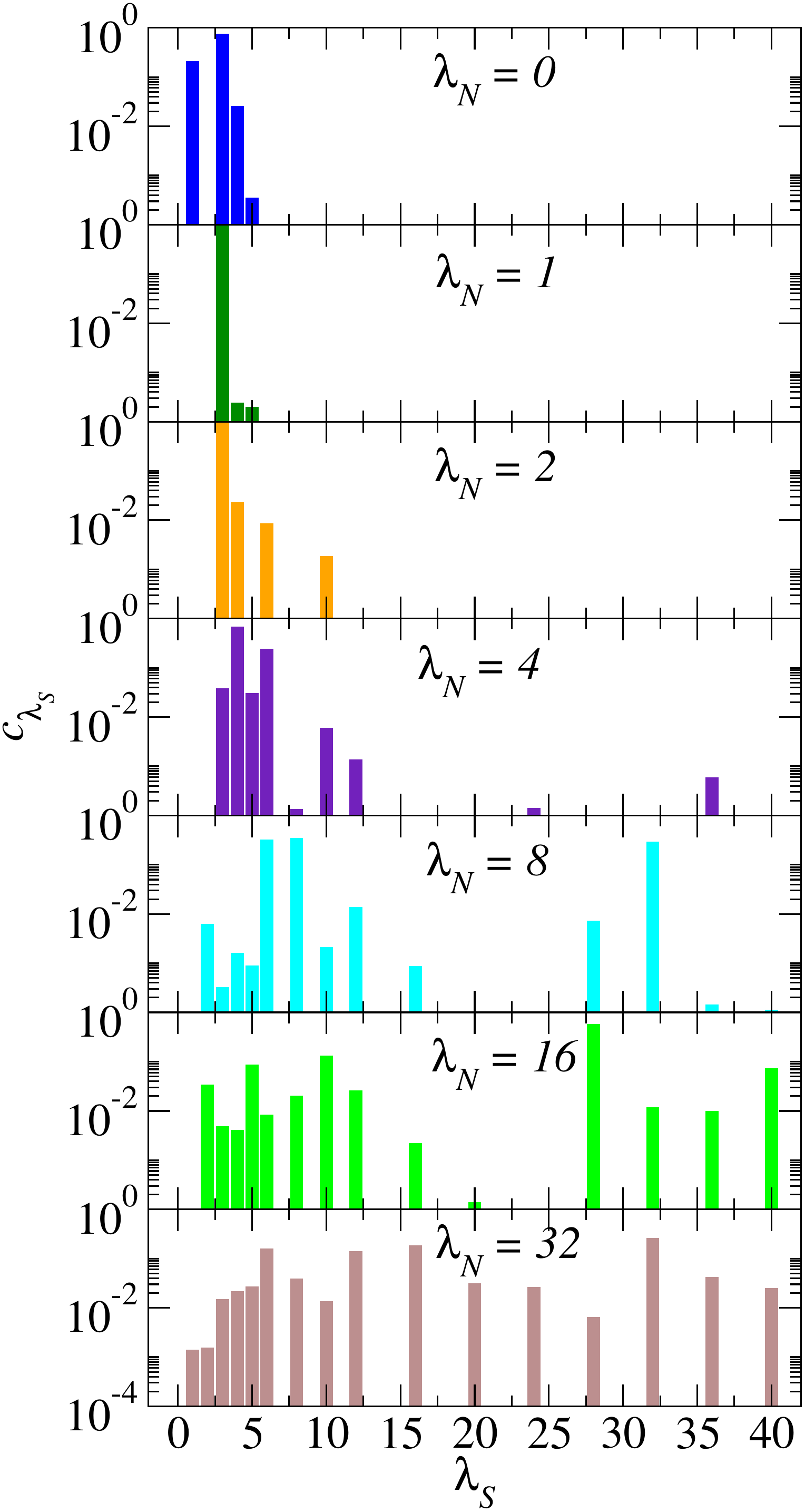}
\caption{Relative concentrations of the ionic surfactant-like patches, $\{c_{\lambda_S}\}$ resulting from the transfer learning procedure, at the indicated Nafion hydrations, $\lambda_N$. These parameters quantify the mapping between the two materials in terms of Eq.~\ref{eq:map}. The data are discussed in depth in the main text.}
\label{fig:weights}
\end{figure}

Also, while $\lambda_N$ unambiguously fixes the total {\em macroscopic} hydration level, it is a dubious measure of the {\em local} content of water, that can be inhomogeneously distributed in the ionomer. We conclude that it is appropriate to allow for an additional degree of freedom, admitting ionic surfactants-like patches at different values of $\lambda_S$ to coexist locally in a Nafion configuration at $\lambda_N$. We therefore seek for a mapping in the form,
\begin{equation}
\Sigma_N(\lambda_N)\rightarrow\;\underset{\lambda_S}{\bigcup}\;\Sigma_S(\lambda_S).
\label{eq:map}
\end{equation}
In all generality, this must associate to a member of the Nafion configurations set $\Sigma_N$ at hydration $\lambda_N$, a whole set of ionic surfactant-like patches, $\{\Sigma_S\}$, identified by the labels $\{\lambda_S\}$. We also request the procedure to directly provide the relative concentrations of the patches, $\{c_{\lambda_S}\}$, with $0\le c_{\lambda_S}\le 1$ and $\sum_{\lambda_S} c_{\lambda_S}=1$. A dominating $c^*=c_{\lambda^*_S}$, would therefore associate the most-likely structural motif in Nafion to the (known) symmetry of the template ionic surfactant system at $\lambda^*_S$. In contrast, multiple high-valued $\{c_{\lambda_S}\}$ would signal the coexistence of different nano-morphologies.  

To implement the above idea, we have resorted to the notion of {\em transfer learning}~\cite{CS231n2016}, where we consider the CNN with weights trained on a data-set, and use those previously learned features to predict new classes ({\em e.~g.}, from cars to trucks) with a partial re-training of a few CNN layers, if necessary. Here, we have fed the above classifier {\em without} any retraining with the Nafion grid data analogous to Fig.~\ref{fig:voxels} at hydration $\lambda_N$. We have therefore ran inference on Nafion configurations by using a CNN trained on surfactants. Note that inference in general provides an entire set of probabilities $\{p_{\lambda_S}\}$, next associating the detected class to the $\max\;[p_{\lambda_S}]$. We can therefore legitimately pose $c_{\lambda_S}=p_{\lambda_S}$, for each $\lambda_S$.

In Fig.~\ref{fig:weights} we show the $c_{\lambda_S}$ concentrations corresponding to the indicated values of $\lambda_N$. This is, we believe, the central result of this work. A few observations are in order. First, the distributions of $c_{\lambda_S}$ strongly depend on $\lambda_N$ and, especially at intermediate to high $\lambda_N$, no single morphology dominates over the others. This can appear obvious but it also implies that, on this basis, it is not possible to claim the existence of one local morphology typical of Nafion at all hydrations. 

Second, surprisingly the ionic surfactants-like patches are not "populated" in sequence {\color{black}({\em i.~e.}, lamellar/hexagonal/micellar phases)} continuously traversing the associated phase diagram, as one could expect trying to image an increasing quantity of water absorbing into the ionomer. Indeed, for $\lambda_N\le 4$, extended flat local morphologies are the most probable, as expected. At intermediate $\lambda_N$, however, we detect more or less spherical micelles, coexisting with the previous topologically different locally flat structures. Elongated structures intermediate between the two above eventually appear only for $\lambda_N>16$. At this point the ionomer appears to the CNN as a (quite flat) disordered distribution of all possible nanomorphologies. Overall these results show that our approach is indeed able to grasp highly non trivial structural features modifications, and provides a quite precise view of the swelling behavior of the membrane, with implications that we describe in the following.
\section{The static structure factors}
\label{sect:sk}
The above findings are totally grounded on volumetric real space data. We now ask if they allow to reach any conclusion based on the static structure factors,
\begin{equation}
S(Q)=\frac{1}{N}\left<\sum_{i,j} b_i b_j e^{i\; \vec{Q}\cdot(\vec{R}_i-\vec{R}_j)}\right>_{|\vec{Q}|=Q}.
\label{eq:sk-def}   
\end{equation}
Here, $\vec{R}_{i, j}$ and $b_{i, j}$ are the position vectors and the scattering lengths of beads $i$ and $j$, respectively, $N$ is the total number of beads, and $\langle \rangle$  indicates both the thermodynamic and the spherical average over wave vectors $\vec{Q}$ of modulus $Q$. We now attempt to determine if a decomposition in the vein of Eq.~(\ref{eq:map}) is still meaningful for this succinct structural descriptor. We therefore express the ionomer $S_N(Q;\lambda_N)$ as a weighted superposition of those of the surfactants,
\begin{equation}
S_N(Q;\lambda_N)\simeq\sum_{\lambda_S} c_{\lambda_S} S_S(Q;\lambda_S),
\label{eq:sk-reconstructed}
\end{equation}
where the $c_{\lambda_S}$ are those of Fig.~\ref{fig:weights}.
\begin{figure}[t]
\centering
\includegraphics[width=0.49\textwidth]{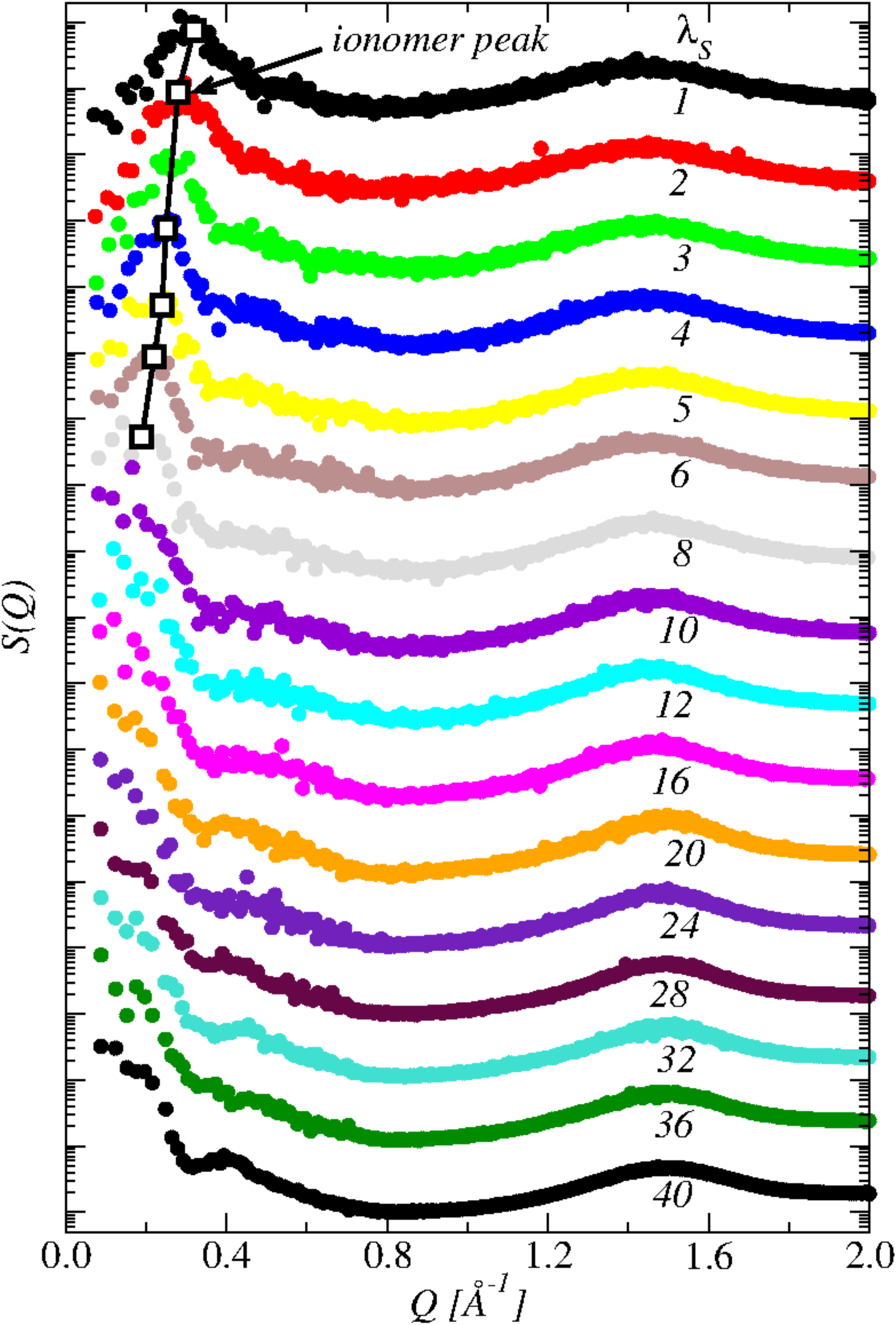}
\caption{Surfactants static structure factors, $S(Q)$, at the indicated values of hydration, $\lambda_S$. {\color{black}We show the position of the ionomer peak with the open white squares for $\lambda_S<10$}. Note that the data have been arbitrarily shifted vertically, to avoid overlaps. These data are used to estimate the Nafion structure factors, as detailed in the text.}
\label{fig:sk-surfactants}
\end{figure}
\begin{figure}[t]
\centering
\includegraphics[width=0.41\textwidth]{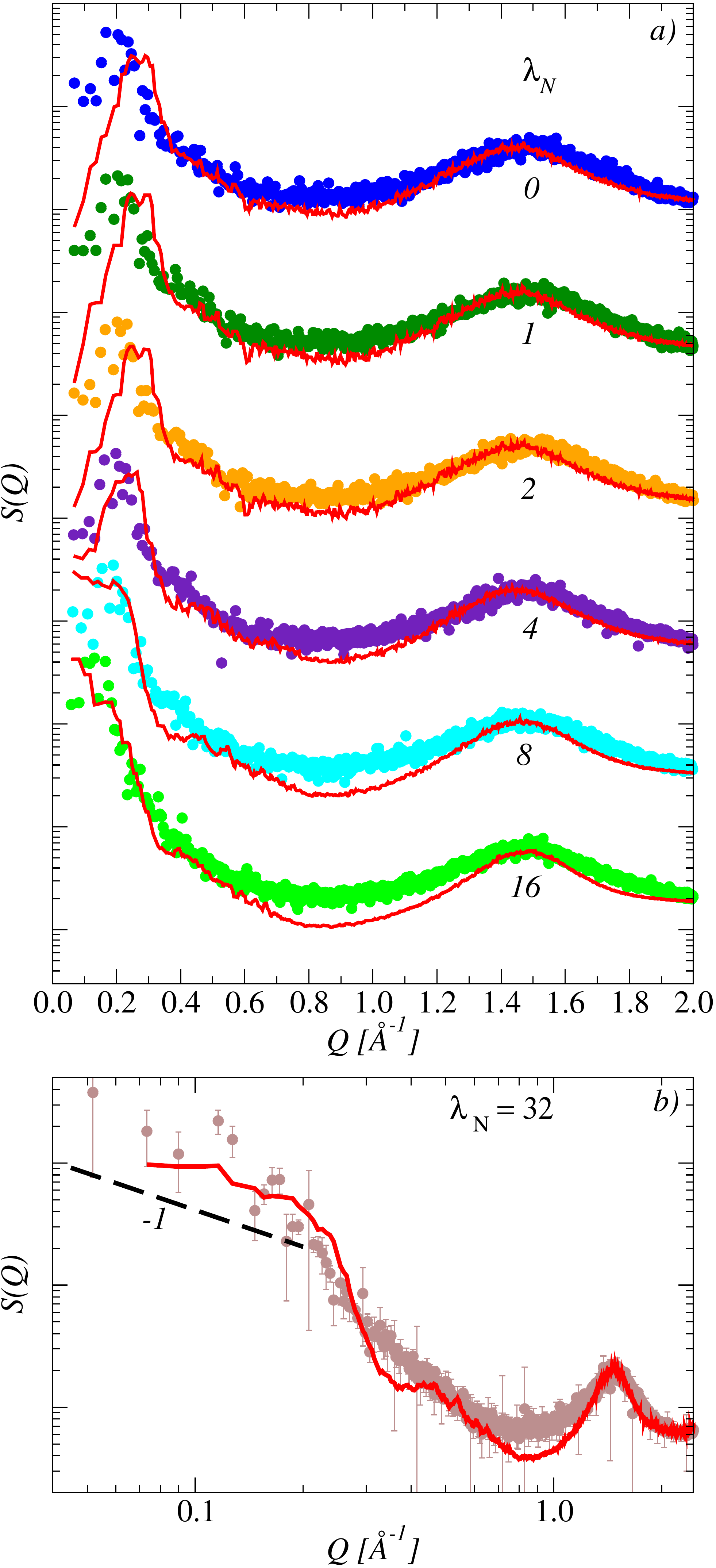}
\caption{a) Expressing the Nafion structure factors as weighted superpositions of the ionic surfactant structure factors at multiple $\lambda_S$, according to Eq.~(\ref{eq:sk-reconstructed}). {\color{black}Closed symbols are the $S_N(Q)$ at the indicated values of $\lambda_N$ calculated directly from the ionomer MD configurations, while the red curves are the reconstructed structure factors.} b) Details of the reconstructed $S_N(Q)$ at $\lambda=32$. The dashed line is a guide-for-the eye indicated the $\propto Q^{-1}$ behavior expected for the matrix knee signal at very low $Q$. These data can be compared qualitatively to those reported in Fig.~3 of~\cite{gebel2005neutron}.}
\label{fig:sk-nafion}
\end{figure}

We have used the ionic surfactants data to calculate the $S_S(Q;\lambda_S)$ at all values of $\lambda_S$, and we show our results in Fig.~\ref{fig:sk-surfactants}. (The data have been shifted arbitrarily to avoid overlaps.) As expected, at the lowest values of $\lambda_S$ they show the typical ionomer peak (indicated by the open squares), corresponding to length scales $\simeq 21$~\AA~\cite{hanot2015water}, but without any sign of high-order Bragg peaks signaling long-range lamellar order, as discussed above. On increasing  $\lambda_S$, the ionomer peak position shifts to lower values indicating swelling with increasing water content, {\em i.e.}, an increase of the average size of the ionic domains~\cite{hanot2015water,Berrod2017}. {\color{black} Note that for larger values of $\lambda_S$ this feature is not visible any longer. This is due to the fact that, at these hydration values, swelling pushes the ionomer peak position to very low values of $Q$, which are unreachable with sufficient resolution by employing the present simulation box sizes, significantly smaller than those considered  in~\cite{hanot2015water}.}

By following the same procedure we have also calculated the $S_N(Q;\lambda_N)$ for the ionomer, at all investigated values of $\lambda_N$, that we show with closed symbols in Fig.~\ref{fig:sk-nafion}(a) and (b). (These data have also been arbitrarily shifted to avoid overlaps.) Here, again, a well defined ionomer peak at around $Q\simeq 0.2$~\AA\; survives for $\lambda_N<16$, eventually completely disappearing in the phase separated systems at higher values of hydration.

We have next directly inserted in the r.h.s of Eq.~(\ref{eq:sk-reconstructed}) the structure factors of Fig.~\ref{fig:sk-surfactants} and the $c_{\lambda_S}$  of Fig.~\ref{fig:weights}. We plot the result of this procedure in Fig.~\ref{fig:sk-nafion}~(a) with the red solid lines. Note that, although they have been shifted together with the corresponding ionomer $S(Q)$ of exactly the same amount to avoid overlaps, no additional adjustment whatsoever has been performed on the data. Also, we performed MD simulations in the $(NPT)$-ensemble, which implies important modifications of the system density with $\lambda_{N,S}$, different for the two materials. We did not try to scale out this difference from our data neither. This procedure must therefore be considered as a crude agnostic reconstruction of the ionomer structure factor, without any adjustable parameter, and by no means a fitting procedure. 

On this basis, the reconstruction in Fig.~\ref{fig:sk-nafion}~(a) is in fact quite accurate, with a very good match of the data at large $Q$, as expected at small length scales, and a systematic underestimation of the intensity at intermediate values. At small $Q$-values the match is not perfect at low $\lambda_N$, with slightly overestimated ionomer peak positions, while the general shape of the peak is satisfactorily recovered. At the highest $\lambda_S$, in contrast, the diverging phase separation region is described with great accuracy. In Fig.~\ref{fig:sk-nafion}~b) we show our results at the highest available $\lambda_S=32$ (on a double logarithmic scale), together with the (guide-for-the-eye) expected small-$Q$ behavior, $\propto Q^{-1}$, which is indeed quite convincingly obeyed. Note that this picture is similar to what measured in neutrons scattering experiments, as one can realize by comparison with Fig.~3 of~\cite{gebel2005neutron}, with the sequence of matrix knee, ionomer peak and WAXS peak, from low to high $Q$.
\section{Discussion and conclusions}
\label{sect:discussion}
In this work we have applied advances in machine learning, 3-dimensional convolutional neural networks, to address the Nafion multi-scale morphology transformations upon variation of the absorbed water content. With a fresh view of a venerable but still debated issue, we have proposed to describe the structure of the ionomer as a collection of patches corresponding to well characterized local morphologies of related ionic surfactants systems. 

We have shown that a convincing mapping can be established between the two classes of materials, without resorting to any prior for the underlying structural model. We have first trained a CNN to classify a vast set of surfactants configurations based on their water content. Next, the trained CNN has been applied in exactly the same form to Nafion instances in a range of hydrations. We have demonstrated that this procedure directly provides an estimate of the relative concentrations of the different patches, together with their variation with the water content of the ionomer. We have finally exploited the latter data to reproduce the Nafion static structure factor in terms of those of the ionic surfactants. This work underlines a few facts, that we discuss below.
\begin{figure*}[t]
\centering
\includegraphics[width=0.85\textwidth]{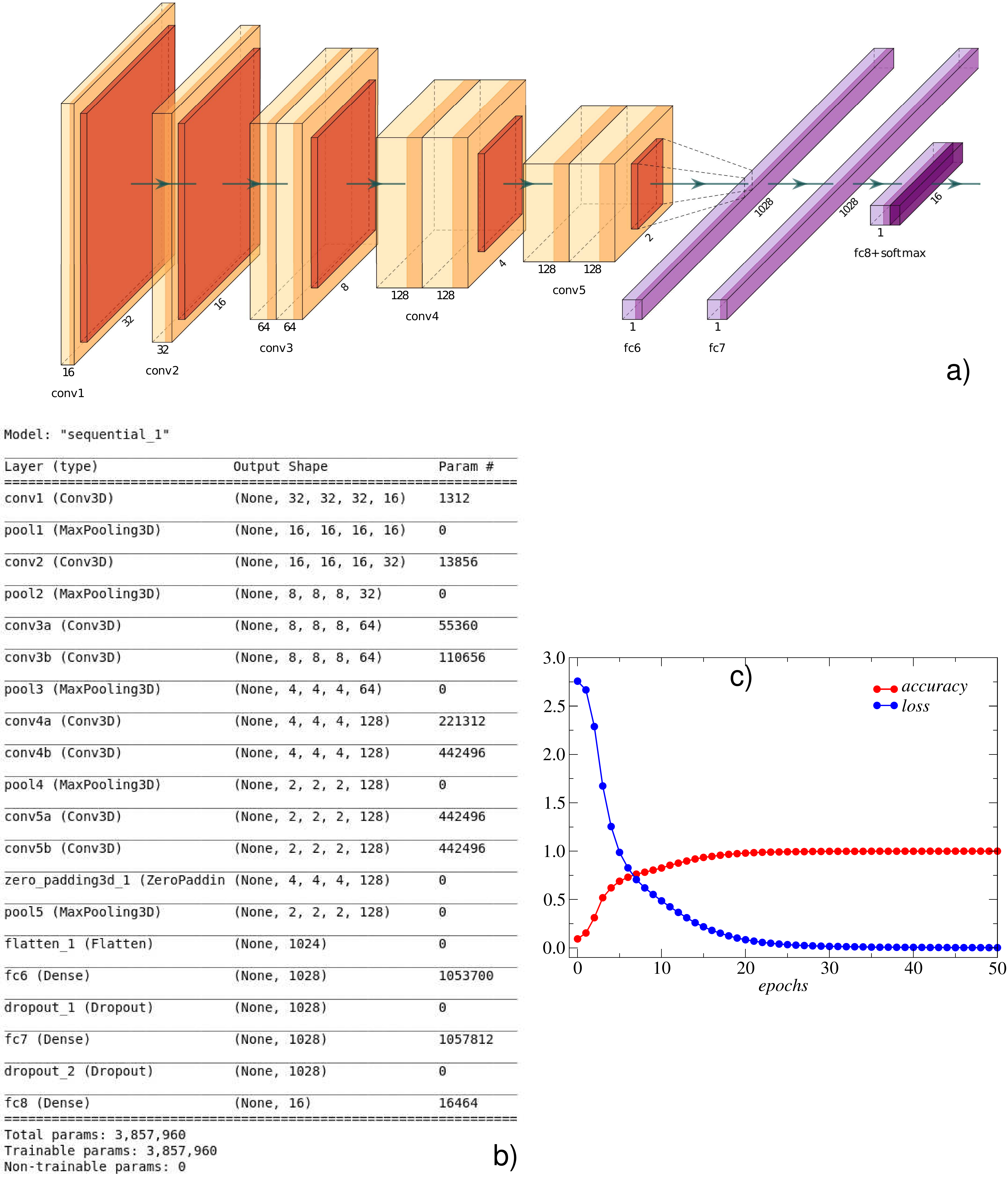} 
\caption{{\em a) and b)} The 3-dimensional convolutional neural network architecture used in this work. General architecture, details of the layers, and total number of trainable parameters are included. {\em c)} Loss and accuracy evolution with the number of epochs during the training stage. The curves are evaluated by using the validation data-set and demonstrate the efficiency of the CNN implementation.}
\label{fig:cnn}
\end{figure*}

{\color{black} The observation that {\em "there is nothing like "the" morphology of Nafion"} contained in ~\cite{kreuer2013critical} is supported by Fig.~\ref{fig:weights}. This indeed shows that the distribution of local ionic surfactants patches providing the highest similarity with the Nafion morphology widely changes with $\lambda_N$. The pretension to identify a single stylized geometric model which should be a good descriptor of the ionomer nano-morphology at any water content is therefore significantly weakened by our findings. On the other hand, we have observed that extended flat, spherical and cylindrical/elongated phases pile up, emerging sequentially upon increasing hydration, which is {\em not} the order they appear on the surfactants phase diagram. This implies that, consistent with the macroscopic/microscopic swelling measurements of~\cite{elliott2000interpretation}, the latter cannot be interpreted as a continuous process of affine deformations of the phase separated domains, also due to the coexistence of multiple local arrangements of completely different topology. In addition, note that in~\cite{hanot2015water} we demonstrated that already in pure surfactant phases the main features of the interfaces ({\em i.~e.}, curvature) change significantly with the water content, modifying in highly non-trivial ways the local wetting properties which remain, however, spatially homogeneous. In the ionomer, in contrast, patches with completely different wetting behavior coexist, determining a strongly inhomogeneous surface tension distribution. How the nature of interactions and the presence of structural disorder contribute to build the situation depicted in Fig.~\ref{fig:weights} is an open crucial question.}

In addition, we are aware of the fact that an ionomer of the complexity of Nafion is intrinsically an out-of-equilibrium system, with a long-range structure evolving on time scales much larger than those explored here. Our results therefore refer to the morphology of well stabilized instances of our model ionomer, on time scales of the order of a few tenths of nanoseconds. We cannot exclude the occurrence of a substantially different picture for much longer aging times.

We have also shown how the proposed mapping can be exploited to express the Nafion static structure factor in terms of those of the ionic surfactants at multiple water contents, without any adjustable parameter. We believe that this possibility is not obvious, when one realizes that the highly structured volumetric data of Fig.~\ref{fig:voxels}, and the extremely succinct scalar observables of Fig.~\ref{fig:sk-surfactants} and Figs.~\ref{fig:sk-nafion}~a) and~b), although obviously strongly related, still contain information of quite different nature. This opens exciting perspectives, especially in the direction of system properties, beyond structure. It would be extremely interesting, for instance, to verify if a similar treatment could provide new insight on the correlated dynamics ({\em e.g.}, diffusion coefficients or relaxation times at different length scales~\cite{hanot2016sub,Berrod2017}) of water and ions in the two materials. These are quantities effectively measured by Quasi-Elastic Neutrons Scattering or Nuclear Magnetic Resonance, which would substantially expand the reach of this work which, as a matter of fact, is entirely based on numerical data. It would be beneficial to clarify how experimental data could be integrated into a similar framework.

We conclude by observing that in our treatment, for both training and inference, we have only focused on the CNN {\em outputs}. The network functionalities we have exploited, however, rest on an {\em internal} representation of the data-sets, which is multi-scale in nature going deeper into the layered CNN architecture. A statistical analysis of the internal state of the network could therefore clarify what the network actually {\em sees}~\cite{qin2018convolutional}. {\color{black}For instance, one can identify what sort of input maximizes the filters included in each layer of the trained CNN~\cite{chollet2016convolutional}, providing a direct visualization of the hierarchical decomposition of the CNN “visual space”. By analyzing these new data, {\em e.~g.} determining spatial correlations of voxels, one should be able to extract typical length scales associated to the size of the ionic channels, or even to more complex quantities, like tortuosity. All-together, these methods could therefore allow us to extract important information that go well beyond the classification task we discussed here, opening new perspectives in the understanding of such complex soft materials.}
\section{Methods}
\label{sect:methods}
\subsection{Molecular Dynamics simulation}
\label{subsect:MD-methods}
All the details of molecular structures and force fields used for the sulfonated ionic surfactants and the Nafion systems can be found in~\cite{hanot2015water} and~\cite{Berrod2017}, respectively. Very briefly, we have employed a united-atoms representation for the surfactant macro-molecule and the side chain of Nafion similar to the model of~\cite{allahyarov2009simulation}. {\color{black}In this description, the hydrophobic uncharged section is represented with a series of 7 neutral beads, each representing an entire $CF_2$ group (3 atoms). It is attached to the head group, schematized by two charged beads, for the sulfur atom (1 atom) and for the $O_3$ group (3 atoms), respectively, with a total charge $q=-e$. In the ionomer, the side chains are next regularly grafted along the flexible highly hydrophobic polymer backbone, also formed by $CF_2$ beads. We generated polymers with 14 $CF_2$ monomers between each side-chain, and 100 side-chains per polymer. This spacing between side-chains results in an equivalent weight of $1080\;g/eq$, a value that is close to the commonly studied Nafion 212 whose equivalent weight is $1100\;g/eq$.} Beads which are not directly bonded interact with Lennard Jones and Coulombic potentials. The latter are truncated and screened according to a modified version of the damped shifted force model~\cite{fennell2006ewald}. 

In contrast, we have considered the atomic resolution three-points rigid SPC/E model for water molecules~\cite{berendsen1987missing}, slightly modified to include the above truncation method. Charge neutrality is imposed by adding one hydronium ion per charged head group, represented by the (four-points) model of~\cite{kusaka1998binary}. The hydration level $\lambda$ is fixed by tuning the ratio of the number of water molecules over that of the ions. In the case of surfactants, we have fixed a total number of interacting units $N\simeq 32\times 10^3$ and chosen the number of each species according to $\lambda$. For Nafion, in contrast, we have fixed the number of polymers and chosen the number of absorbed molecules according to $\lambda$, which amounts to values of $N$ in the range $29\times 10^3$ to $133\times 10^3$. {\color{black}As a reference, the system with $\lambda=0$ ($32$) contains $29160$ ($132840$) interacting units which, considering the definition of the coarse-grained beads detailed above, correspond to $76680$ ($180360$) atoms in total, in a simulation box of a linear size of $9.2$ ($12.2$)~nm.}

The production runs following thermalization span for both materials a total time scale of 20~ns. In the case of surfactants, we have repeated the procedure twice, starting from completely independent configurations at high temperature. Along the trajectories we have appropriately dumped both complete system configurations, used for the calculations of $S(Q)$ via Eq.~(\ref{eq:sk-def}), and the (averaged on-the-fly) voxels samples of Fig.~\ref{fig:voxels}, which we have employed for training and inference with our CNN, as discussed in the main text and below. All technical details about the MD runs performed with LAMMPS~\cite{plimpton1995fast} are the same as reported in~\cite{hanot2015water,hanot2016sub,Berrod2017}.
\subsection{The Convolutional Neural Network}
\label{sub_sect:neural_networks}
3-dimensional convolutional neural networks like, for instance, VoxNet of~\cite{maturana2015voxnet} are employed for analysis of data where the temporal or volumetric context is important. The ability to analyze a series of frames or images has led to the use of 3-dim CNN for many applications, ranging from evaluation of medical imaging (see, for instance, an application to  segmentation in~\cite{casamitjana20163d}) to action recognition~\cite{ji20123d}, among others. While in the first case the volumetric nature of data is obvious, in the latter case the process of analyzing the position of objects in a sequence of 2-dim images, like a video, leads to $(2+1)$-dimensional information.

Our CNN architecture is adapted with a few variations from the C3D network of~\cite{tran2015learning}, which belongs to this last class. It is a spatio-temporal feature learning 3-dimensional convolutional network for video data-set applications, where we have substituted the time dimension with a third spatial dimension~\footnote{C3D has been originally trained on the Sports-1M dataset~\cite{karpathy2014large}, at the time the largest video classification benchmark.}. Details of our implementation are presented in Fig.~\ref{fig:cnn}~a) and b). 

The network is constituted by 8 convolution layers, 5 pooling layers, followed by two fully connected layers, and the final softmax output layer, for a total of $3,857,960$ trainable parameters (see Fig.~\ref{fig:cnn}~a) and b)). All 3-dim convolution kernels are $3\times 3 \times 3$ with stride 1 in all dimensions, and padding leaving unaltered the size of the feature maps. The variable number of filters is indicated in Fig.~\ref{fig:cnn}~b), and ranges from 16 to 128 going deeper in the structure. All (max-)pooling kernels are $2\times 2 \times 2$ with stride 2. Each fully connected layer has 1028 output units. The rectified linear unit (ReLu) activation function is employed everywhere. 

{\color{black}The final training set was constituted by 25600 voxel configurations of size $32^3$ with 3 channels, calculated from the ionic surfactants instances as described above. Note that, due to the fixed size of the grid, the size of the employed voxel and therefore the available spatial resolution, depends on hydration. Considering the typical simulation box lengths indicated above and the size (3) of the employed convolutional kernels, the minimum spatial resolution is in the range $7$ to $9$~\AA.}

Training was performed by using the Adam optimizer, with a mini-batch size of 32, while the learning rate was divided by 2 in the case of no loss improvement after 2 epochs. The optimization was terminated after 50 epochs. All hyper-parameters values have been optimized by trial-and-error. 
The training stage is depicted in Fig.~\ref{fig:cnn}~c), where we plot both accuracy and loss calculated on the validation set, as a function of the number of epochs. Loss and accuracy were evaluated on a validation set containing 3200 configurations at evenly distributed values of $\lambda_S$. {\color{red} In order to exclude the possibility of overfitting, we checked that the plot of the validation loss decreases to a constant value with a small gap with the training loss, and that no sign of subsequent increase of the validation loss for a larger number of epochs was observed.}

Inference was applied to other 3200 configurations, obviously not included in the training set. Inference on Nafion voxels configurations were performed analogously, on a test set of 7000 samples for 7 values of the hydration $\lambda_N$.  

For the implementation we have employed standard Python libraries, including the Keras~\cite{chollet2015keras} high-level API to the TensorFlow~\cite{abadi2019tensorflow} machine leaning framework back-end, or the HDF5 high performance data software library~\cite{hdf5}. Training and inference have been executed on a self-designed workstation, including a Ryzen Threadripper 1950X (3.4 GHz, 32 threads) CPU, 64 GB of DDR4-3000 memory, and two 3~GHz GTX 1080Ti GPUs. 
\begin{acknowledgments}
L. Dumortier acknowledges ILL/ESRF funding in the framework of the X-Ray and Neutron Science -- International Student Summer Program at ILL/ESRF (2019). S. Mossa is supported by ANR-19-CE06/0025 (MOVEYOURION). S. Mossa thanks S. Hanot for useful discussions on advanced statistical treatments of tomography data.
\end{acknowledgments}
\bibliography{references}

\begin{thebibliography}{40}
\expandafter\ifx\csname natexlab\endcsname\relax\def\natexlab#1{#1}\fi
\expandafter\ifx\csname bibnamefont\endcsname\relax
  \def\bibnamefont#1{#1}\fi
\expandafter\ifx\csname bibfnamefont\endcsname\relax
  \def\bibfnamefont#1{#1}\fi
\expandafter\ifx\csname citenamefont\endcsname\relax
  \def\citenamefont#1{#1}\fi
\expandafter\ifx\csname url\endcsname\relax
  \def\url#1{\texttt{#1}}\fi
\expandafter\ifx\csname urlprefix\endcsname\relax\def\urlprefix{URL }\fi
\providecommand{\bibinfo}[2]{#2}
\providecommand{\eprint}[2][]{\url{#2}}

\bibitem[{\citenamefont{Kreuer}(2013)}]{kreuer2013ion}
\bibinfo{author}{\bibfnamefont{K.-D.} \bibnamefont{Kreuer}},
  \bibinfo{journal}{Chem. Mater.} \textbf{\bibinfo{volume}{26}},
  \bibinfo{pages}{361} (\bibinfo{year}{2013}), ISSN \bibinfo{issn}{0897-4756,
  1520-5002}, \urlprefix\url{https://doi.org/10.1021/cm402742u}.

\bibitem[{\citenamefont{Gebel}(2000)}]{gebel2000structural}
\bibinfo{author}{\bibfnamefont{G.}~\bibnamefont{Gebel}},
  \bibinfo{journal}{Polymer} \textbf{\bibinfo{volume}{41}},
  \bibinfo{pages}{5829} (\bibinfo{year}{2000}), ISSN \bibinfo{issn}{0032-3861},
  \urlprefix\url{https://doi.org/10.1016/s0032-3861(99)00770-3}.

\bibitem[{\citenamefont{Kreuer and Portale}(2013)}]{kreuer2013critical}
\bibinfo{author}{\bibfnamefont{K.-D.} \bibnamefont{Kreuer}} \bibnamefont{and}
  \bibinfo{author}{\bibfnamefont{G.}~\bibnamefont{Portale}},
  \bibinfo{journal}{Adv. Funct. Mater.} \textbf{\bibinfo{volume}{23}},
  \bibinfo{pages}{5390} (\bibinfo{year}{2013}), ISSN \bibinfo{issn}{1616-301X},
  \urlprefix\url{https://doi.org/10.1002/adfm.201300376}.

\bibitem[{\citenamefont{Hsu and Gierke}(1983)}]{hsu1983ion}
\bibinfo{author}{\bibfnamefont{W.~Y.} \bibnamefont{Hsu}} \bibnamefont{and}
  \bibinfo{author}{\bibfnamefont{T.~D.} \bibnamefont{Gierke}},
  \bibinfo{journal}{J. Membr. Sci.} \textbf{\bibinfo{volume}{13}},
  \bibinfo{pages}{307} (\bibinfo{year}{1983}), ISSN \bibinfo{issn}{0376-7388},
  \urlprefix\url{https://doi.org/10.1016/s0376-7388(00)81563-x}.

\bibitem[{\citenamefont{Rubatat et~al.}(2004)\citenamefont{Rubatat, Gebel, and
  Diat}}]{rubatat2004fibrillar}
\bibinfo{author}{\bibfnamefont{L.}~\bibnamefont{Rubatat}},
  \bibinfo{author}{\bibfnamefont{G.}~\bibnamefont{Gebel}}, \bibnamefont{and}
  \bibinfo{author}{\bibfnamefont{O.}~\bibnamefont{Diat}},
  \bibinfo{journal}{Macromolecules} \textbf{\bibinfo{volume}{37}},
  \bibinfo{pages}{7772} (\bibinfo{year}{2004}), ISSN \bibinfo{issn}{0024-9297,
  1520-5835}, \urlprefix\url{https://doi.org/10.1021/ma049683j}.

\bibitem[{\citenamefont{Schmidt-Rohr and Chen}(2007)}]{schmidt2008parallel}
\bibinfo{author}{\bibfnamefont{K.}~\bibnamefont{Schmidt-Rohr}}
  \bibnamefont{and} \bibinfo{author}{\bibfnamefont{Q.}~\bibnamefont{Chen}},
  \bibinfo{journal}{Nature Mater} \textbf{\bibinfo{volume}{7}},
  \bibinfo{pages}{75} (\bibinfo{year}{2007}), ISSN \bibinfo{issn}{1476-1122,
  1476-4660}, \urlprefix\url{https://doi.org/10.1038/nmat2074}.

\bibitem[{\citenamefont{Venkatnathan et~al.}(2007)\citenamefont{Venkatnathan,
  Devanathan, and Dupuis}}]{venkatnathan2007atomistic}
\bibinfo{author}{\bibfnamefont{A.}~\bibnamefont{Venkatnathan}},
  \bibinfo{author}{\bibfnamefont{R.}~\bibnamefont{Devanathan}},
  \bibnamefont{and} \bibinfo{author}{\bibfnamefont{M.}~\bibnamefont{Dupuis}},
  \bibinfo{journal}{J. Phys. Chem. B} \textbf{\bibinfo{volume}{111}},
  \bibinfo{pages}{7234} (\bibinfo{year}{2007}), ISSN \bibinfo{issn}{1520-6106,
  1520-5207}, \urlprefix\url{https://doi.org/10.1021/jp0700276}.

\bibitem[{\citenamefont{Allahyarov and
  Taylor}(2011)}]{allahyarov2011simulation}
\bibinfo{author}{\bibfnamefont{E.}~\bibnamefont{Allahyarov}} \bibnamefont{and}
  \bibinfo{author}{\bibfnamefont{P.~L.} \bibnamefont{Taylor}},
  \bibinfo{journal}{J. Polym. Sci. B Polym. Phys.}
  \textbf{\bibinfo{volume}{49}}, \bibinfo{pages}{368} (\bibinfo{year}{2011}),
  ISSN \bibinfo{issn}{0887-6266},
  \urlprefix\url{https://doi.org/10.1002/polb.22191}.

\bibitem[{\citenamefont{Vishnyakov and Neimark}(2014)}]{vishnyakov2014self}
\bibinfo{author}{\bibfnamefont{A.}~\bibnamefont{Vishnyakov}} \bibnamefont{and}
  \bibinfo{author}{\bibfnamefont{A.~V.} \bibnamefont{Neimark}},
  \bibinfo{journal}{J. Phys. Chem. B} \textbf{\bibinfo{volume}{118}},
  \bibinfo{pages}{11353} (\bibinfo{year}{2014}), ISSN \bibinfo{issn}{1520-6106,
  1520-5207}, \urlprefix\url{https://doi.org/10.1021/jp504975u}.

\bibitem[{\citenamefont{Knox and Voth}(2010)}]{knox2010probing}
\bibinfo{author}{\bibfnamefont{C.~K.} \bibnamefont{Knox}} \bibnamefont{and}
  \bibinfo{author}{\bibfnamefont{G.~A.} \bibnamefont{Voth}},
  \bibinfo{journal}{J. Phys. Chem. B} \textbf{\bibinfo{volume}{114}},
  \bibinfo{pages}{3205} (\bibinfo{year}{2010}), ISSN \bibinfo{issn}{1520-6106,
  1520-5207}, \urlprefix\url{https://doi.org/10.1021/jp9112409}.

\bibitem[{\citenamefont{Elliott et~al.}(2011)\citenamefont{Elliott, Wu,
  Paddison, and Moore}}]{elliott2011unified}
\bibinfo{author}{\bibfnamefont{J.~A.} \bibnamefont{Elliott}},
  \bibinfo{author}{\bibfnamefont{D.}~\bibnamefont{Wu}},
  \bibinfo{author}{\bibfnamefont{S.~J.} \bibnamefont{Paddison}},
  \bibnamefont{and} \bibinfo{author}{\bibfnamefont{R.~B.} \bibnamefont{Moore}},
  \bibinfo{journal}{Soft Matter} \textbf{\bibinfo{volume}{7}},
  \bibinfo{pages}{6820} (\bibinfo{year}{2011}), ISSN \bibinfo{issn}{1744-683X,
  1744-6848}, \urlprefix\url{https://doi.org/10.1039/c1sm00002k}.

\bibitem[{\citenamefont{Allen et~al.}(2014)\citenamefont{Allen, Comolli,
  Kusoglu, Modestino, Minor, and Weber}}]{allen2015morphology}
\bibinfo{author}{\bibfnamefont{F.~I.} \bibnamefont{Allen}},
  \bibinfo{author}{\bibfnamefont{L.~R.} \bibnamefont{Comolli}},
  \bibinfo{author}{\bibfnamefont{A.}~\bibnamefont{Kusoglu}},
  \bibinfo{author}{\bibfnamefont{M.~A.} \bibnamefont{Modestino}},
  \bibinfo{author}{\bibfnamefont{A.~M.} \bibnamefont{Minor}}, \bibnamefont{and}
  \bibinfo{author}{\bibfnamefont{A.~Z.} \bibnamefont{Weber}},
  \bibinfo{journal}{ACS Macro Lett.} \textbf{\bibinfo{volume}{4}},
  \bibinfo{pages}{1} (\bibinfo{year}{2014}), ISSN \bibinfo{issn}{2161-1653,
  2161-1653}, \urlprefix\url{https://doi.org/10.1021/mz500606h}.

\bibitem[{\citenamefont{Lyonnard et~al.}(2010)\citenamefont{Lyonnard, Berrod,
  Brüning, Gebel, Guillermo, Ftouni, Ollivier, and
  Frick}}]{lyonnard2010perfluorinated}
\bibinfo{author}{\bibfnamefont{S.}~\bibnamefont{Lyonnard}},
  \bibinfo{author}{\bibfnamefont{Q.}~\bibnamefont{Berrod}},
  \bibinfo{author}{\bibfnamefont{B.-A.} \bibnamefont{Brüning}},
  \bibinfo{author}{\bibfnamefont{G.}~\bibnamefont{Gebel}},
  \bibinfo{author}{\bibfnamefont{A.}~\bibnamefont{Guillermo}},
  \bibinfo{author}{\bibfnamefont{H.}~\bibnamefont{Ftouni}},
  \bibinfo{author}{\bibfnamefont{J.}~\bibnamefont{Ollivier}}, \bibnamefont{and}
  \bibinfo{author}{\bibfnamefont{B.}~\bibnamefont{Frick}},
  \bibinfo{journal}{Eur. Phys. J. Spec. Top.} \textbf{\bibinfo{volume}{189}},
  \bibinfo{pages}{205} (\bibinfo{year}{2010}), ISSN \bibinfo{issn}{1951-6355,
  1951-6401}, \urlprefix\url{https://doi.org/10.1140/epjst/e2010-01324-x}.

\bibitem[{\citenamefont{Hanot et~al.}(2015)\citenamefont{Hanot, Lyonnard, and
  Mossa}}]{hanot2015water}
\bibinfo{author}{\bibfnamefont{S.}~\bibnamefont{Hanot}},
  \bibinfo{author}{\bibfnamefont{S.}~\bibnamefont{Lyonnard}}, \bibnamefont{and}
  \bibinfo{author}{\bibfnamefont{S.}~\bibnamefont{Mossa}},
  \bibinfo{journal}{Soft Matter} \textbf{\bibinfo{volume}{11}},
  \bibinfo{pages}{2469} (\bibinfo{year}{2015}), ISSN \bibinfo{issn}{1744-683X,
  1744-6848}, \urlprefix\url{https://doi.org/10.1039/c5sm00179j}.

\bibitem[{\citenamefont{Hanot et~al.}(2016)\citenamefont{Hanot, Lyonnard, and
  Mossa}}]{hanot2016sub}
\bibinfo{author}{\bibfnamefont{S.}~\bibnamefont{Hanot}},
  \bibinfo{author}{\bibfnamefont{S.}~\bibnamefont{Lyonnard}}, \bibnamefont{and}
  \bibinfo{author}{\bibfnamefont{S.}~\bibnamefont{Mossa}},
  \bibinfo{journal}{Nanoscale} \textbf{\bibinfo{volume}{8}},
  \bibinfo{pages}{3314} (\bibinfo{year}{2016}), ISSN \bibinfo{issn}{2040-3364,
  2040-3372}, \urlprefix\url{https://doi.org/10.1039/c5nr05853h}.

\bibitem[{\citenamefont{Berrod et~al.}(2017)\citenamefont{Berrod, Hanot,
  Guillermo, Mossa, and Lyonnard}}]{Berrod2017}
\bibinfo{author}{\bibfnamefont{Q.}~\bibnamefont{Berrod}},
  \bibinfo{author}{\bibfnamefont{S.}~\bibnamefont{Hanot}},
  \bibinfo{author}{\bibfnamefont{A.}~\bibnamefont{Guillermo}},
  \bibinfo{author}{\bibfnamefont{S.}~\bibnamefont{Mossa}}, \bibnamefont{and}
  \bibinfo{author}{\bibfnamefont{S.}~\bibnamefont{Lyonnard}},
  \bibinfo{journal}{Sci Rep} \textbf{\bibinfo{volume}{7}}
  (\bibinfo{year}{2017}), ISSN \bibinfo{issn}{2045-2322},
  \urlprefix\url{https://doi.org/10.1038/s41598-017-08746-9}.

\bibitem[{\citenamefont{LeCun et~al.}(2015)\citenamefont{LeCun, Bengio, and
  Hinton}}]{lecun2015deep}
\bibinfo{author}{\bibfnamefont{Y.}~\bibnamefont{LeCun}},
  \bibinfo{author}{\bibfnamefont{Y.}~\bibnamefont{Bengio}}, \bibnamefont{and}
  \bibinfo{author}{\bibfnamefont{G.}~\bibnamefont{Hinton}},
  \bibinfo{journal}{Nature} \textbf{\bibinfo{volume}{521}},
  \bibinfo{pages}{436} (\bibinfo{year}{2015}), ISSN \bibinfo{issn}{0028-0836,
  1476-4687}, \urlprefix\url{https://doi.org/10.1038/nature14539}.

\bibitem[{\citenamefont{Rickman et~al.}(2019)\citenamefont{Rickman, Lookman,
  and Kalinin}}]{rickman2019materials}
\bibinfo{author}{\bibfnamefont{J.}~\bibnamefont{Rickman}},
  \bibinfo{author}{\bibfnamefont{T.}~\bibnamefont{Lookman}}, \bibnamefont{and}
  \bibinfo{author}{\bibfnamefont{S.}~\bibnamefont{Kalinin}},
  \bibinfo{journal}{Acta Mater.} \textbf{\bibinfo{volume}{168}},
  \bibinfo{pages}{473} (\bibinfo{year}{2019}), ISSN \bibinfo{issn}{1359-6454},
  \urlprefix\url{https://doi.org/10.1016/j.actamat.2019.01.051}.

\bibitem[{\citenamefont{Carleo et~al.}(2019)\citenamefont{Carleo, Cirac,
  Cranmer, Daudet, Schuld, Tishby, Vogt-Maranto, and
  Zdeborová}}]{carleo2019machine}
\bibinfo{author}{\bibfnamefont{G.}~\bibnamefont{Carleo}},
  \bibinfo{author}{\bibfnamefont{I.}~\bibnamefont{Cirac}},
  \bibinfo{author}{\bibfnamefont{K.}~\bibnamefont{Cranmer}},
  \bibinfo{author}{\bibfnamefont{L.}~\bibnamefont{Daudet}},
  \bibinfo{author}{\bibfnamefont{M.}~\bibnamefont{Schuld}},
  \bibinfo{author}{\bibfnamefont{N.}~\bibnamefont{Tishby}},
  \bibinfo{author}{\bibfnamefont{L.}~\bibnamefont{Vogt-Maranto}},
  \bibnamefont{and}
  \bibinfo{author}{\bibfnamefont{L.}~\bibnamefont{Zdeborová}},
  \bibinfo{journal}{Rev. Mod. Phys.} \textbf{\bibinfo{volume}{91}},
  \bibinfo{pages}{045002} (\bibinfo{year}{2019}), ISSN
  \bibinfo{issn}{0034-6861, 1539-0756},
  \urlprefix\url{https://doi.org/10.1103/revmodphys.91.045002}.

\bibitem[{\citenamefont{Ferguson}(2017)}]{ferguson2017machine}
\bibinfo{author}{\bibfnamefont{A.~L.} \bibnamefont{Ferguson}},
  \bibinfo{journal}{J. Phys.: Condens. Matter} \textbf{\bibinfo{volume}{30}},
  \bibinfo{pages}{043002} (\bibinfo{year}{2017}), ISSN
  \bibinfo{issn}{0953-8984, 1361-648X},
  \urlprefix\url{https://doi.org/10.1088/1361-648x/aa98bd}.

\bibitem[{\citenamefont{Kalinin et~al.}(2015)\citenamefont{Kalinin, Sumpter,
  and Archibald}}]{kalinin2015big}
\bibinfo{author}{\bibfnamefont{S.~V.} \bibnamefont{Kalinin}},
  \bibinfo{author}{\bibfnamefont{B.~G.} \bibnamefont{Sumpter}},
  \bibnamefont{and} \bibinfo{author}{\bibfnamefont{R.~K.}
  \bibnamefont{Archibald}}, \bibinfo{journal}{Nature Mater}
  \textbf{\bibinfo{volume}{14}}, \bibinfo{pages}{973} (\bibinfo{year}{2015}),
  ISSN \bibinfo{issn}{1476-1122, 1476-4660},
  \urlprefix\url{https://doi.org/10.1038/nmat4395}.

\bibitem[{\citenamefont{Vasudevan et~al.}(2018)\citenamefont{Vasudevan,
  Laanait, Ferragut, Wang, Geohegan, Xiao, Ziatdinov, Jesse, Dyck, and
  Kalinin}}]{vasudevan2018mapping}
\bibinfo{author}{\bibfnamefont{R.~K.} \bibnamefont{Vasudevan}},
  \bibinfo{author}{\bibfnamefont{N.}~\bibnamefont{Laanait}},
  \bibinfo{author}{\bibfnamefont{E.~M.} \bibnamefont{Ferragut}},
  \bibinfo{author}{\bibfnamefont{K.}~\bibnamefont{Wang}},
  \bibinfo{author}{\bibfnamefont{D.~B.} \bibnamefont{Geohegan}},
  \bibinfo{author}{\bibfnamefont{K.}~\bibnamefont{Xiao}},
  \bibinfo{author}{\bibfnamefont{M.}~\bibnamefont{Ziatdinov}},
  \bibinfo{author}{\bibfnamefont{S.}~\bibnamefont{Jesse}},
  \bibinfo{author}{\bibfnamefont{O.}~\bibnamefont{Dyck}}, \bibnamefont{and}
  \bibinfo{author}{\bibfnamefont{S.~V.} \bibnamefont{Kalinin}},
  \bibinfo{journal}{npj Comput Mater} \textbf{\bibinfo{volume}{4}},
  \bibinfo{pages}{30} (\bibinfo{year}{2018}), ISSN \bibinfo{issn}{2057-3960},
  \urlprefix\url{https://doi.org/10.1038/s41524-018-0086-7}.

\bibitem[{\citenamefont{Plimpton}(1995)}]{plimpton1995fast}
\bibinfo{author}{\bibfnamefont{S.}~\bibnamefont{Plimpton}},
  \bibinfo{journal}{J. Comput. Phys.} \textbf{\bibinfo{volume}{117}},
  \bibinfo{pages}{1} (\bibinfo{year}{1995}), ISSN \bibinfo{issn}{0021-9991},
  \urlprefix\url{https://doi.org/10.1006/jcph.1995.1039}.

\bibitem[{\citenamefont{Chollet}(2015)}]{chollet2015keras}
\bibinfo{author}{\bibfnamefont{F.}~\bibnamefont{Chollet}},
  \emph{\bibinfo{title}{Keras}}, \bibinfo{howpublished}{\url
  {https://github.com/fchollet/keras}} (\bibinfo{year}{2015}).

\bibitem[{\citenamefont{Abadi et~al.}(2019)\citenamefont{Abadi, Agarwal,
  Barham, Brevdo, Chen, Citro, Corrado, Davis, Dean, Devin
  et~al.}}]{abadi2019tensorflow}
\bibinfo{author}{\bibfnamefont{M.}~\bibnamefont{Abadi}},
  \bibinfo{author}{\bibfnamefont{A.}~\bibnamefont{Agarwal}},
  \bibinfo{author}{\bibfnamefont{P.}~\bibnamefont{Barham}},
  \bibinfo{author}{\bibfnamefont{E.}~\bibnamefont{Brevdo}},
  \bibinfo{author}{\bibfnamefont{Z.}~\bibnamefont{Chen}},
  \bibinfo{author}{\bibfnamefont{C.}~\bibnamefont{Citro}},
  \bibinfo{author}{\bibfnamefont{G.~S.} \bibnamefont{Corrado}},
  \bibinfo{author}{\bibfnamefont{A.}~\bibnamefont{Davis}},
  \bibinfo{author}{\bibfnamefont{J.}~\bibnamefont{Dean}},
  \bibinfo{author}{\bibfnamefont{M.}~\bibnamefont{Devin}},
  \bibnamefont{et~al.}, \bibinfo{journal}{arXiv preprint arXiv:1603.04467}
  (\bibinfo{year}{2019}).

\bibitem[{\citenamefont{Li et~al.}(2016)\citenamefont{Li, Karpathy, and
  Johnson}}]{CS231n2016}
\bibinfo{author}{\bibfnamefont{F.-F.} \bibnamefont{Li}},
  \bibinfo{author}{\bibfnamefont{A.}~\bibnamefont{Karpathy}}, \bibnamefont{and}
  \bibinfo{author}{\bibfnamefont{J.}~\bibnamefont{Johnson}}
  (\bibinfo{year}{2016}), \urlprefix\url{http://cs231n.stanford.edu/}.

\bibitem[{\citenamefont{Gebel and Diat}(2005)}]{gebel2005neutron}
\bibinfo{author}{\bibfnamefont{G.}~\bibnamefont{Gebel}} \bibnamefont{and}
  \bibinfo{author}{\bibfnamefont{O.}~\bibnamefont{Diat}},
  \bibinfo{journal}{Fuel Cells} \textbf{\bibinfo{volume}{5}},
  \bibinfo{pages}{261} (\bibinfo{year}{2005}), ISSN \bibinfo{issn}{1615-6846,
  1615-6854}, \urlprefix\url{https://doi.org/10.1002/fuce.200400080}.

\bibitem[{\citenamefont{Elliott et~al.}(2000)\citenamefont{Elliott, Hanna,
  Elliott, and Cooley}}]{elliott2000interpretation}
\bibinfo{author}{\bibfnamefont{J.}~\bibnamefont{Elliott}},
  \bibinfo{author}{\bibfnamefont{S.}~\bibnamefont{Hanna}},
  \bibinfo{author}{\bibfnamefont{A.}~\bibnamefont{Elliott}}, \bibnamefont{and}
  \bibinfo{author}{\bibfnamefont{G.}~\bibnamefont{Cooley}},
  \bibinfo{journal}{Macromolecules} \textbf{\bibinfo{volume}{33}},
  \bibinfo{pages}{4161} (\bibinfo{year}{2000}), ISSN \bibinfo{issn}{0024-9297,
  1520-5835}, \urlprefix\url{https://doi.org/10.1021/ma991113+}.

\bibitem[{\citenamefont{Qin et~al.}(2018)\citenamefont{Qin, Yu, Liu, and
  Chen}}]{qin2018convolutional}
\bibinfo{author}{\bibfnamefont{Z.}~\bibnamefont{Qin}},
  \bibinfo{author}{\bibfnamefont{F.}~\bibnamefont{Yu}},
  \bibinfo{author}{\bibfnamefont{C.}~\bibnamefont{Liu}}, \bibnamefont{and}
  \bibinfo{author}{\bibfnamefont{X.}~\bibnamefont{Chen}},
  \bibinfo{journal}{Mathematical Foundations of Computing}
  \textbf{\bibinfo{volume}{1}}, \bibinfo{pages}{149} (\bibinfo{year}{2018}),
  ISSN \bibinfo{issn}{2577-8838},
  \urlprefix\url{https://doi.org/10.3934/mfc.2018008}.

\bibitem[{\citenamefont{Chollet}(2016)}]{chollet2016convolutional}
\bibinfo{author}{\bibfnamefont{F.}~\bibnamefont{Chollet}},
  \bibinfo{journal}{The Keras Blog} \textbf{\bibinfo{volume}{30}}
  (\bibinfo{year}{2016}).

\bibitem[{\citenamefont{Allahyarov and
  Taylor}(2009)}]{allahyarov2009simulation}
\bibinfo{author}{\bibfnamefont{E.}~\bibnamefont{Allahyarov}} \bibnamefont{and}
  \bibinfo{author}{\bibfnamefont{P.~L.} \bibnamefont{Taylor}},
  \bibinfo{journal}{J. Phys. Chem. B} \textbf{\bibinfo{volume}{113}},
  \bibinfo{pages}{610} (\bibinfo{year}{2009}), ISSN \bibinfo{issn}{1520-6106,
  1520-5207}, \urlprefix\url{https://doi.org/10.1021/jp8047746}.

\bibitem[{\citenamefont{Fennell and Gezelter}(2006)}]{fennell2006ewald}
\bibinfo{author}{\bibfnamefont{C.~J.} \bibnamefont{Fennell}} \bibnamefont{and}
  \bibinfo{author}{\bibfnamefont{J.~D.} \bibnamefont{Gezelter}},
  \bibinfo{journal}{J. Chem. Phys.} \textbf{\bibinfo{volume}{124}},
  \bibinfo{pages}{234104} (\bibinfo{year}{2006}), ISSN
  \bibinfo{issn}{0021-9606, 1089-7690},
  \urlprefix\url{https://doi.org/10.1063/1.2206581}.

\bibitem[{\citenamefont{Berendsen et~al.}(1987)\citenamefont{Berendsen,
  Grigera, and Straatsma}}]{berendsen1987missing}
\bibinfo{author}{\bibfnamefont{H.}~\bibnamefont{Berendsen}},
  \bibinfo{author}{\bibfnamefont{J.}~\bibnamefont{Grigera}}, \bibnamefont{and}
  \bibinfo{author}{\bibfnamefont{T.}~\bibnamefont{Straatsma}},
  \bibinfo{journal}{J. Phys. Chem.} \textbf{\bibinfo{volume}{91}},
  \bibinfo{pages}{6269} (\bibinfo{year}{1987}), ISSN \bibinfo{issn}{0022-3654,
  1541-5740}, \urlprefix\url{https://doi.org/10.1021/j100308a038}.

\bibitem[{\citenamefont{Kusaka et~al.}(1998)\citenamefont{Kusaka, Wang, and
  Seinfeld}}]{kusaka1998binary}
\bibinfo{author}{\bibfnamefont{I.}~\bibnamefont{Kusaka}},
  \bibinfo{author}{\bibfnamefont{Z.-G.} \bibnamefont{Wang}}, \bibnamefont{and}
  \bibinfo{author}{\bibfnamefont{J.}~\bibnamefont{Seinfeld}},
  \bibinfo{journal}{J. Chem. Phys.} \textbf{\bibinfo{volume}{108}},
  \bibinfo{pages}{6829} (\bibinfo{year}{1998}), ISSN \bibinfo{issn}{0021-9606,
  1089-7690}, \urlprefix\url{https://doi.org/10.1063/1.476097}.

\bibitem[{\citenamefont{Maturana and Scherer}(2015)}]{maturana2015voxnet}
\bibinfo{author}{\bibfnamefont{D.}~\bibnamefont{Maturana}} \bibnamefont{and}
  \bibinfo{author}{\bibfnamefont{S.}~\bibnamefont{Scherer}}, in
  \emph{\bibinfo{booktitle}{2015 IEEE/RSJ International Conference on
  Intelligent Robots and Systems (IROS)}}, \bibinfo{organization}{IEEE}
  (\bibinfo{publisher}{IEEE}, \bibinfo{year}{2015}), pp.
  \bibinfo{pages}{922--928}, ISBN \bibinfo{isbn}{9781479999941},
  \urlprefix\url{https://doi.org/10.1109/iros.2015.7353481}.

\bibitem[{\citenamefont{Casamitjana et~al.}(2016)\citenamefont{Casamitjana,
  Puch, Aduriz, and Vilaplana}}]{casamitjana20163d}
\bibinfo{author}{\bibfnamefont{A.}~\bibnamefont{Casamitjana}},
  \bibinfo{author}{\bibfnamefont{S.}~\bibnamefont{Puch}},
  \bibinfo{author}{\bibfnamefont{A.}~\bibnamefont{Aduriz}}, \bibnamefont{and}
  \bibinfo{author}{\bibfnamefont{V.}~\bibnamefont{Vilaplana}}, in
  \emph{\bibinfo{booktitle}{International Workshop on Brainlesion: Glioma,
  Multiple Sclerosis, Stroke and Traumatic Brain Injuries}}
  (\bibinfo{organization}{Springer}, \bibinfo{year}{2016}), pp.
  \bibinfo{pages}{150--161}.

\bibitem[{\citenamefont{Ji et~al.}(2013)\citenamefont{Ji, Xu, Yang, and
  Yu}}]{ji20123d}
\bibinfo{author}{\bibfnamefont{S.}~\bibnamefont{Ji}},
  \bibinfo{author}{\bibfnamefont{W.}~\bibnamefont{Xu}},
  \bibinfo{author}{\bibfnamefont{M.}~\bibnamefont{Yang}}, \bibnamefont{and}
  \bibinfo{author}{\bibfnamefont{K.}~\bibnamefont{Yu}}, \bibinfo{journal}{IEEE
  Trans. Pattern Anal. Mach. Intell.} \textbf{\bibinfo{volume}{35}},
  \bibinfo{pages}{221} (\bibinfo{year}{2013}), ISSN \bibinfo{issn}{0162-8828,
  2160-9292}, \urlprefix\url{https://doi.org/10.1109/tpami.2012.59}.

\bibitem[{\citenamefont{Tran et~al.}(2015)\citenamefont{Tran, Bourdev, Fergus,
  Torresani, and Paluri}}]{tran2015learning}
\bibinfo{author}{\bibfnamefont{D.}~\bibnamefont{Tran}},
  \bibinfo{author}{\bibfnamefont{L.}~\bibnamefont{Bourdev}},
  \bibinfo{author}{\bibfnamefont{R.}~\bibnamefont{Fergus}},
  \bibinfo{author}{\bibfnamefont{L.}~\bibnamefont{Torresani}},
  \bibnamefont{and} \bibinfo{author}{\bibfnamefont{M.}~\bibnamefont{Paluri}},
  in \emph{\bibinfo{booktitle}{2015 IEEE International Conference on Computer
  Vision (ICCV)}} (\bibinfo{publisher}{IEEE}, \bibinfo{year}{2015}), pp.
  \bibinfo{pages}{4489--4497}, ISBN \bibinfo{isbn}{9781467383912},
  \urlprefix\url{https://doi.org/10.1109/iccv.2015.510}.

\bibitem[{\citenamefont{Koranne}(2010)}]{hdf5}
\bibinfo{author}{\bibfnamefont{S.}~\bibnamefont{Koranne}},
  \emph{\bibinfo{title}{Hierarchical data format 5: HDF5}}
  (\bibinfo{publisher}{Springer US}, \bibinfo{year}{2010}), chap.
  \bibinfo{chapter}{Hierarchical Data Format 5 : HDF5}, pp.
  \bibinfo{pages}{191--200}, ISBN \bibinfo{isbn}{9781441977182, 9781441977199},
  \urlprefix\url{https://doi.org/10.1007/978-1-4419-7719-9_10}.

\bibitem[{\citenamefont{Karpathy et~al.}(2014)\citenamefont{Karpathy, Toderici,
  Shetty, Leung, Sukthankar, and Fei-Fei}}]{karpathy2014large}
\bibinfo{author}{\bibfnamefont{A.}~\bibnamefont{Karpathy}},
  \bibinfo{author}{\bibfnamefont{G.}~\bibnamefont{Toderici}},
  \bibinfo{author}{\bibfnamefont{S.}~\bibnamefont{Shetty}},
  \bibinfo{author}{\bibfnamefont{T.}~\bibnamefont{Leung}},
  \bibinfo{author}{\bibfnamefont{R.}~\bibnamefont{Sukthankar}},
  \bibnamefont{and} \bibinfo{author}{\bibfnamefont{L.}~\bibnamefont{Fei-Fei}},
  in \emph{\bibinfo{booktitle}{2014 IEEE Conference on Computer Vision and
  Pattern Recognition}} (\bibinfo{publisher}{IEEE}, \bibinfo{year}{2014}), pp.
  \bibinfo{pages}{1725--1732}, ISBN \bibinfo{isbn}{9781479951185},
  \urlprefix\url{https://doi.org/10.1109/cvpr.2014.223}.

\end{thebibliography}
\end{document}